\documentclass[reprint,floatfix,superscriptaddress]{revtex4-2}

\usepackage{amsmath}
\usepackage{amssymb}
\usepackage{soul}
\usepackage{ulem}
\usepackage{graphicx}
\usepackage{dcolumn}
\usepackage{bm}
\usepackage{siunitx}
\DeclareSIUnit{\bits}{bits}
\usepackage[T1]{fontenc}
\usepackage{mathptmx}
\usepackage{makecell}
\usepackage{threeparttable}
\usepackage{nicefrac}
\usepackage{hyperref}
\usepackage{braket}
\usepackage{wasysym}
\usepackage{tikz}
\usepackage{romannum}
\usepackage{color}
\definecolor{hotblue}{RGB}{46,48,146}
\hypersetup{
            colorlinks=true,
            linkcolor=hotblue,
            citecolor=hotblue,
            urlcolor=hotblue,
            }
\definecolor{blue}{RGB}{97,111,183}
\definecolor{orange}{RGB}{239,134,54}
\definecolor{green}{RGB}{0,133,125}
\definecolor{red}{RGB}{235,127,126}
\definecolor{yellow}{RGB}{235,235,100}

\begin{document}

\title{High-rate intercity quantum key distribution with a semiconductor single-photon source}

\author{Jingzhong Yang}
\affiliation{Institut f{\"u}r Festk{\"o}rperphysik, Leibniz Universit{\"a}t Hannover, Appelstra{\ss}e~2, 30167~Hannover, Germany}

\author{Zenghui Jiang}
\affiliation{Institut f{\"u}r Festk{\"o}rperphysik, Leibniz Universit{\"a}t Hannover, Appelstra{\ss}e~2, 30167~Hannover, Germany}

\author{Frederik Benthin}
\affiliation{Institut f{\"u}r Festk{\"o}rperphysik, Leibniz Universit{\"a}t Hannover, Appelstra{\ss}e~2, 30167~Hannover, Germany}

\author{Joscha Hanel}
\affiliation{Institut f{\"u}r Festk{\"o}rperphysik, Leibniz Universit{\"a}t Hannover, Appelstra{\ss}e~2, 30167~Hannover, Germany}

\author{Tom Fandrich}
\affiliation{Institut f{\"u}r Festk{\"o}rperphysik, Leibniz Universit{\"a}t Hannover, Appelstra{\ss}e~2, 30167~Hannover, Germany}

\author{Raphael Joos}
\affiliation{Institut f{\"u}r Halbleiteroptik und Funktionelle Grenzfl{\"a}chen, Center for Integrated Quantum Science and Technology (IQ\textsuperscript{ST}) and SCoPE, University of Stuttgart, Allmandring 3, 70569 Stuttgart, Germany}

\author{Stephanie Bauer}
\affiliation{Institut f{\"u}r Halbleiteroptik und Funktionelle Grenzfl{\"a}chen, Center for Integrated Quantum Science and Technology (IQ\textsuperscript{ST}) and SCoPE, University of Stuttgart, Allmandring 3, 70569 Stuttgart, Germany}

\author{Sascha Kolatschek}
\affiliation{Institut f{\"u}r Halbleiteroptik und Funktionelle Grenzfl{\"a}chen, Center for Integrated Quantum Science and Technology (IQ\textsuperscript{ST}) and SCoPE, University of Stuttgart, Allmandring 3, 70569 Stuttgart, Germany}

\author{Ali Hreibi}
\affiliation{Physikalisch-Technische Bundesanstalt, Bundesallee 100, 38116 Braunschweig, Germany}

\author{Eddy Patrick Rugeramigabo}
\affiliation{Institut f{\"u}r Festk{\"o}rperphysik, Leibniz Universit{\"a}t Hannover, Appelstra{\ss}e~2, 30167~Hannover, Germany}

\author{Michael Jetter}
\affiliation{Institut f{\"u}r Halbleiteroptik und Funktionelle Grenzfl{\"a}chen, Center for Integrated Quantum Science and Technology (IQ\textsuperscript{ST}) and SCoPE, University of Stuttgart, Allmandring 3, 70569 Stuttgart, Germany}

\author{Simone Luca Portalupi}
\affiliation{Institut f{\"u}r Halbleiteroptik und Funktionelle Grenzfl{\"a}chen, Center for Integrated Quantum Science and Technology (IQ\textsuperscript{ST}) and SCoPE, University of Stuttgart, Allmandring 3, 70569 Stuttgart, Germany}

\author{Michael Zopf}
\affiliation{Institut f{\"u}r Festk{\"o}rperphysik, Leibniz Universit{\"a}t Hannover, Appelstra{\ss}e~2, 30167~Hannover, Germany}
\affiliation{Laboratorium f{\"u}r Nano- und Quantenengineering, Leibniz Universit{\"a}t Hannover, Schneiderberg 39, 30167 Hannover, Germany}

\author{Peter Michler}
\affiliation{Institut f{\"u}r Halbleiteroptik und Funktionelle Grenzfl{\"a}chen, Center for Integrated Quantum Science and Technology (IQ\textsuperscript{ST}) and SCoPE, University of Stuttgart, Allmandring 3, 70569 Stuttgart, Germany}

\author{Stefan K{\"u}ck}
\affiliation{Physikalisch-Technische Bundesanstalt, Bundesallee 100, 38116 Braunschweig, Germany}

\author{Fei Ding}
\email{fei.ding@fkp.uni-hannover.de}

\affiliation{Institut f{\"u}r Festk{\"o}rperphysik, Leibniz Universit{\"a}t Hannover, Appelstra{\ss}e~2, 30167~Hannover, Germany}
\affiliation{Laboratorium f{\"u}r Nano- und Quantenengineering, Leibniz Universit{\"a}t Hannover, Schneiderberg 39, 30167 Hannover, Germany}

\begin{abstract}
{Quantum key distribution (QKD) enables the transmission of information that is secure against general attacks by eavesdroppers. The use of on-demand quantum light sources in QKD protocols is expected to help improve security and maximum tolerable loss. Semiconductor quantum dots (QDs) are a promising building block for quantum communication applications because of the deterministic emission of single photons with high brightness and low multiphoton contribution. Here we report on the first intercity QKD experiment using a bright deterministic single photon source. A BB84 protocol based on polarisation encoding is realised using the high-rate single photons in the telecommunication C-band emitted from a semiconductor QD embedded in a circular Bragg grating structure. Utilising the \SI{79}{\km} long link with \SI{25.49}{\dB} loss (equivalent to \SI{130}{\km} for the direct-connected optical fibre) between the German cities of Hannover and Braunschweig, a record-high secret key bits per pulse of \SI{4.8e-5}{} with an average quantum bit error ratio of $\sim$\SI{0.65}{\percent} are demonstrated. An asymptotic maximum tolerable loss of \SI{28.11}{\dB} is found, corresponding to a length of \SI{144}{\km} of standard telecommunication fibre. {Deterministic semiconductor sources therefore challenge state-of-the-art QKD protocols} and have the potential to excel in measurement device independent protocols and quantum repeater applications.}
\end{abstract}

\maketitle
\section{Introduction}



Realms of communication that transcend the limitations of traditional networks can be accessed by establishing a `quantum internet' \cite{Wehner2018,Gyongyosi2022} through the distribution of quantum light states. Sharing quantum bits of information with distant nodes via optical fibre or free space (satellite) enables new applications such as quantum teleportation \cite{Bennett1993,Boschi1998,Bouwmeester1997}, quantum cloud computing \cite{Soeparno2021,Taleb2020} or quantum sensor networks \cite{Qian2021,Ge2018}. A primary advantage of quantum communication lies in its ability to ensure unambiguous security for modern communication networks, a security that is increasingly threatened by the rapid advancement of quantum computing technologies \cite{Aaronson2017,Preskill2018,Zhong2020}. Hence, Quantum Key Distribution (QKD) has attracted worldwide attention for its unique ability to provide security based on the principles of quantum mechanics \cite{Lo2014}, surpassing the capabilities of classical cryptography \cite{Rivest1978}.

The QKD landscape has evolved significantly over the years, using a variety of protocols and spanning fibre networks \cite{Chen2021,Liu2023} and satellite-to-ground free-space links \cite{Bedington2017,Lu2022}. Despite this progress, the establishment of large networks currently requires the use of intermediate 'trusted nodes' \cite{Scarani2009}, which provide limited security that can only be fully restored by the implementation of quantum repeaters \cite{Briegel1998}. Furthermore, conventional quantum light sources based on weak laser pulses \cite{Xu2020} or spontaneous parametric down-conversion \cite{Schneeloch2019} struggle with a delicate balance between brightness and multi-photon emissions to resist photon number splitting attacks. Decoy state QKD offers a potential solution \cite{Hwang2003}, but at the cost of increased complexity and a penalty in the secret key rate (SKR) \cite{Scarani2009}. 

Semiconductor single photon sources (SPSs) hold immense potential in revolutionising large-scale quantum communication. Semiconductor quantum dots (QDs) are capable of emitting indistinguishable single photons on demand with unprecedented efficiency and purity \cite{Lu2021,Vajner2022}, offering strong advantages for QKD \cite{Vajner2022,Couteau2023}. {In particular, for measurement-device-independent (MDI) QKD {\cite{FerreiradaSilva2013}}, which requires high visibility of Hong-Ou-Mandel interference between two independent single-photon sources, a scheme involving QDs {\cite{Owen2021}} can significantly improve the key rate {\cite{Zhou2018}}.} QDs also offer great prospects for the realisation of quantum repeaters, as they allow for inherent storage of quantum information \cite{Zaporski2023} and can emit photonic cluster states \cite{Schwartz2016}. The success of these QDs in the wavelength range between \SI{780}{\nano\meter} and \SI{900}{\nano\meter} is expected to be built on by the continued development of QDs emitting at the telecom bands. Quantum communication experiments utilising QDs have demonstrated their ability to link university campuses and metropolitan areas \cite{Alleaume2004,Rau2014,Xiang2019,Basset2021,Schimpf2021,Zahidy2023}. However, the lack of bright single-photon signals in the telecommunication bands have hindered progress beyond these boundaries to intercity distances. Nevertheless, a recent breakthrough \cite{Nawrath2023} has enabled the emission of bright single photons with high emission rates thanks to Purcell enhancement, from a QD device directly in the telecommunication C-band and therefore expanding the horizons of quantum communication.

Here we report on the first intercity QKD experiments with a deterministic single-photon source. A semiconductor quantum dot embedded in a circular Bragg grating (CBG) efficiently emitting single photons of high purity in the telecommunication C-band is employed in conjunction with polarisation encoding in the standard BB84 protocol \cite{Bennett2014}. The photons are routed on a \SI{79}{\km} long deployed fibre between the German cities of Hannover and Braunschweig, featuring a loss of \SI{25.49(0.02)}{\dB} corresponding to a standard telecom fibre length of \SI{130.32}{\km}. We verify that high-rate secret key transmission and a low quantum bit error ratio (QBER) of $\sim$\SI{0.65}{\percent} are ensured for \SI{35}{\hour}. An average secret key bits (SKBs) per pulse of more than \SI{2e-5}{} in the finite-key regime can be reached over an acquisition time of \SI{30}{\minute}. Positive key rates are determined achievable for distances up to \SI{144}{\km} corresponding to \SI{28.11}{\dB} loss in the laboratory, highlighting the competitiveness of semiconductor SPSs for quantum communication applications.


\section{Results}
\subsection{Overview of the experiment}
\begin{figure*}[htbp]
    \centering
    \includegraphics[width=\textwidth]{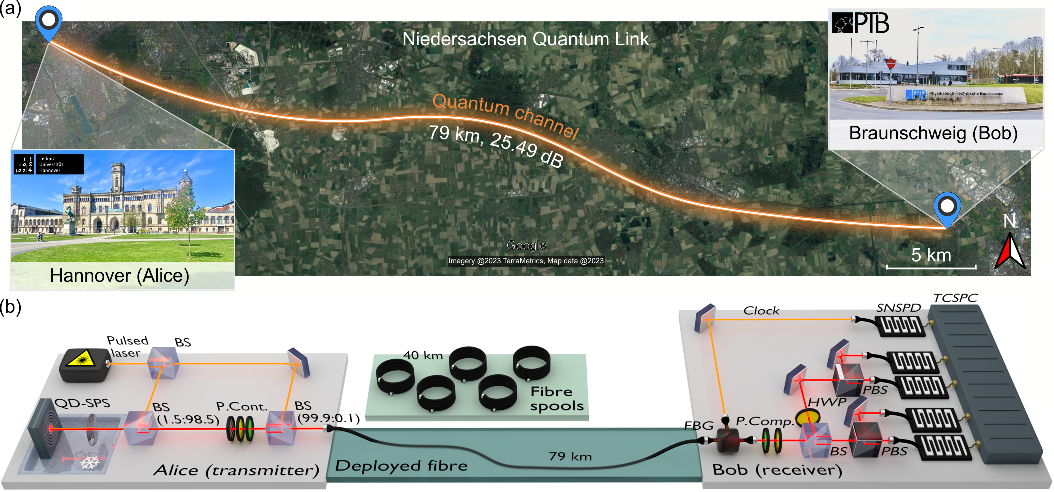}
    \caption{\textbf{Overview of the intercity QKD experiments on the `Niedersachsen quantum link' using single photons from a semiconductor quantum dot (QD).} (a) Distribution of quantum bits between Hannover (Alice) and Braunschweig (Bob) over \SI{79}{\km} of deployed fibre with a total loss of \SI{25.49}{\dB}. Map data from Google (©2023 Google). (b) Sketch of the experimental setup. The QD-based SPS of the transmitter is mounted in a cryostat and excited by a pulsed laser at different clock rates (CRs) (\SI{76}{\MHz}, \SI{228}{\MHz}, \SI{608}{\MHz}, and \SI{1063}{\MHz}). The emitted single photons are collected by an aspherical lens with a numerical aperture of 0.7. State encoding is performed by a polarisation control module (P. Cont.) comprising a polariser, a half-wave plate (HWP) and a quarter-wave plate (QWP). The single-photon and excitation laser signals are then together coupled into either a sequence of fibre spools or the deployed fibre. In the receiver module, a fibre Bragg grating (FBG) demultiplexes the single photon and laser signals by wavelength. An electronically controlled polarisation compensation (P. Comp.) module with QWP and HWP counteracts polarisation fluctuations in the quantum channels by monitoring and minimising the quantum bit error ratio (QBER). A non-polarising 50:50 beam splitter (BS) then acts as a random selector of the decoding basis, with rectilinear projection in the transmitted path using a polarisation beam splitter (PBS), and diagonal projection in the reflected path using a HWP at an angle of \SI{22.5}{\degree} followed by a PBS. The four single-photon signals and the laser signal are detected at superconducting nanowire single-photon detectors (SNSPDs) and the timing events recorded with a time-correlated single-photon counting (TCSPC) unit.}
    \label{fig:optical_setups}
\end{figure*}

The intercity experiment is performed in the German federal state of Niedersachsen, in which a deployed fibre of $\sim \SI{79}{\km}$ length connects the Leibniz University of Hannover (LUH) and Physikalisch-Technische Bundesanstalt (PTB) Braunschweig, as depicted in Fig. \ref{fig:optical_setups}(a). Alice, located at the LUH, statically prepares polarisation-encoded single photons as $\left \{ \ket{H}, \ket{V}, \ket{D}, \ket{A} \right \}$. Bob, located at the PTB, contains a passive polarisation decoder to measure the polarisation states on two balanced conjugate bases. We denote the rectilinear $\left \{\ket{H},\ket{V}\right \}$ and diagonal $\left \{\ket{D},\ket{A}\right \}$ bases  as Z and X bases, respectively.

In the transmitter of Alice, a pulsed laser {(PriTel, InC.)} at a wavelength of $\SI{1529.8}{\nano\meter}$ and with an adjustable clock rate (CR) is employed to excite the {p-shell of the positively charged trion transition of the} InAs/InGaAs/GaAs QD mounted in a \SI{4}{\kelvin} closed-cycle helium gas cryostat [Fig. \ref{fig:optical_setups}(b)]. The QD, embedded in a CBG photonic structure, emits {circularly polarised} single photons at a wavelength of \SI{1555.9}{\nano\meter} with high brightness. The Purcell effect of the CBG cavity reduces the QD's emission lifetime to \SI{592.5(1.8)}{\pico\second} (see details in the Materials and Methods section), theoretically allowing for an increase of the excitation CR up to GHz. {One linear component of the photon emissions from the QD is particularly favoured in brightness due to the asymmetry of the CBG cavity. The super-conducting nanowire detector (SNSPD) detects an average photon count rate of {\SI{3.591(0.003)}{\MHz}} from the transmitter, while the QD is excited at its saturation power under the CR of {\SI{76}{\MHz}} (see also Materials and Methods). The average number of photons per pulse {\cite{Waks2002}} for the linearly polarised fraction of the single photon emission at the first lens is calculated to be {$\left \langle n \right \rangle = \SI{0.138(0.015)}{}$}, taking into account the efficiencies of the transmitter and detector (see Tab. {\ref{tab:parameters}}). It is worth noting that the value of {$\left \langle n \right \rangle$} differs slightly from the reported extraction efficiency {\cite{Nawrath2023}}. This is because the polariser filters out the single photons with linear polarisation that are not favoured by the CBG cavity. Additionally, a raw blinking-corrected {$g^{(2)}(0)$ value of {\SI{2.43(0.02)}{\percent}}} is measured without any data post-processing (see more details in the Materials and Methods section).}

To first study all of the QKD performance for different transmission distances in the lab, encoded single photons are sent through one or multiple standard telecom fibre spools (ITU-T G.652.D) of $\SI{40}{\km}$ length each. The average loss of $l=$\SI{0.1956(0.0026)}{\dB\per\km} per spool is calibrated in the laboratory, taking into account the insertion loss (see Supplementary Information Sec. \Romannum{2}.A.1). So as to then realise the intercity QKD experiment over the deployed fibre, a reference signal for local synchronisation is required. Therefore, the single photon signals are transmitted over the intercity link together with attenuated pulses from the excitation laser. On the receiver side, these two signals are de-multiplexed and the single photon states are decoded. SNSPDs are used for detecting the single photons and the reference laser signal, which thereby provides a timing reference to the single photon detection events.
\begin{figure}[ht]
    \centering
    \includegraphics[width=0.48\textwidth]{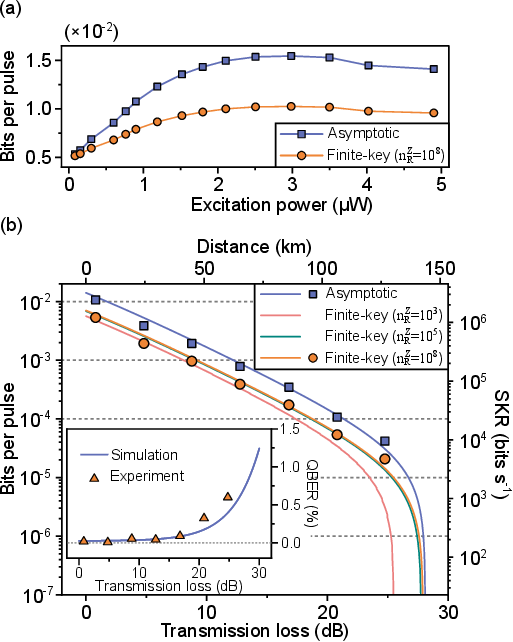}
    \caption{\textbf{In-lab characterisation of the QD based single-photon source and the experimental setup.} Secret key per clock pulse as a function of (a) excitation laser power in the asymptotic and finite-key regime. (b) Secret key bits (SKBs) per pulse as a function of the transmission loss. The maximum tolerable losses (MTLs) for the \textcolor{blue}{asymptotic} case and finite-key block cases with different block sizes of $ n_{R}^{Z}=\textcolor{red}{10^3}, \textcolor{green}{10^5}, \textcolor{orange}{10^8}$ are extracted to be \textcolor{blue}{\SI{28.11}{}}, \textcolor{red}{\SI{25.51}{}}, \textcolor{green}{\SI{27.78}{}} and \textcolor{orange}{\SI{27.95}{}} \SI{}{\dB}, respectively. The inset shows the measured and simulated QBER as a function of transmission loss with the dashed grey line indicating the QBER at zero.}
    \label{fig:lab_qkd}
\end{figure}

One of the figure of merits used to assess the performance of QKD is the SKBs per pulse. In our work, we study this in both the asymptotic and finite-key regimes. For the asymptotic case \cite{Gottesman2004,Cai2009,Waks2002},
\begin{equation}
\begin{aligned}
    S_{A} = p_{sift}  \left \{ \underline{p}_c^{(1)} \left [ 1- h \left ( \overline{e}_1 \right ) \right ] - f_{EC} p_c h \left ( e_{tot}\right ) \right \}
\label{eq:asymp}
\end{aligned}
\end{equation}
where $p_{sift}=p_X^2+(1-p_X)^2$ is the sifting ratio assuming both QKD bases are used for the key generation; $p_X$ is the bias of the projection basis ($p_X = 0.5$ in our case); $\underline{p}_c^{(1)}$ corresponds to the lower bound of detected events for the single-photon state; $h(\cdot)$ is the binary Shannon entropy function; $\overline{e}_1$ denotes the upper bound of the QBER for single-photon states and $e_{tot}$ is the total QBER for all photon number states. For convenience, we assume balanced efficiencies of the receiver ports and SNSPD channels for each polarisation basis. $f_{EC}$ describes the error correction inefficiency of the algorithm, $p_c$ indicates the total detection probability of the photon number states \cite{Bozzio2022,Vyvlecka2023}. 

For the case of a finite block size of the keys, we evaluate the SKBs per pulse using the multiplicative Chernoff bound \cite{Yin2020,Morrison2023},
\begin{equation}
\begin{aligned}
    S_{F} =\frac{\underline{n}_{R,nmp}^{X,Z}}{Rt} \left [  1-h(\overline{\phi}^Z) -\lambda_{EC} - \log_{2}{\frac{2}{\epsilon_{cor}}} - 2\log_{2}{\frac{1}{2\epsilon_{PA}}}\right ]
\label{eq:finite}
\end{aligned}
\end{equation}
Here, $R$ is the CR, $t$ is the acquisition time,  $\underline{n}_{R,nmp}^{X,Z}$ the lower bound of non-multiphoton emissions in the receiver module for X and Z bases, $\overline{\phi}^Z$ the upper bound of the phase error rate, $\lambda_{EC}$ the lower bound of information leakage \cite{Tomamichel2017} and $\epsilon_{cor}$ are the bits used for verification during the error correction process. $\epsilon_{PA}$ is the failure probability of privacy amplification. Table \ref{tab:parameters} presents both the performance of our QD-based SPS and security parameters of our QKD system, in which we consider $\epsilon$-secret $\epsilon_{sec}= 10^{-10}$ and $\epsilon$-correct $\epsilon_{cor} = 10^{-15}$ for reaching $\epsilon_{qkd}$-secure ($\epsilon_{qkd}\ge\epsilon_{sec}+\epsilon_{cor}$) \cite{Bunandar2020}.
\begin{table}[ht]
    \caption{\label{tab:parameters} \textbf{{In-lab QKD system and security parameters.}}}
    \begin{threeparttable}
    \begin{ruledtabular}
    \begin{tabular}{ccc}
    \textrm{\textbf{Description}}&\textrm{\textbf{Parameter}}&\textrm{\makecell[c]{\textbf{Value}}}\\
    \colrule
    Average photon number per pulse & $\left \langle n \right \rangle$  & $ 0.138 $ \\
    Clock rate & $ R $  & $\SI{228}{\MHz}$ \\
    Second-order correlation & $g^{(2)}(0)$  & $ \SI{2.43}{\percent} $ \\
    Transmitter efficiency  & $\eta_{T}$  & $ 0.464 $  \\
    Receiver efficiency  & $\eta_{R}$  & $ 0.740 $  \\
    System misalignment probability &  $p_{mis}$  & $ 2.57 \times 10^{-4} $ \\
    Detector efficiency &  $\eta_{D}$  & $ 0.740 $ \\
    Dark count probability & $p_{dc}$ & $ 8.74\times 10^{-7} $ \\
    Dead time & $\tau_{dt}$ & $ \SI{35.865}{\nano\second} $ \\
    Averaged fibre-spool loss & $l$ & \SI{0.1956}{\dB\per\km} \\
    Field-installed fibre loss & $L$ & \SI{25.49}{\dB} \\
    Parameter estimation failure probability & $\epsilon_{PE}$ & $\nicefrac{2\times 10^{-10}}{3}$ \\
    Error correction failure probability & $\epsilon_{EC}$ & $\nicefrac{10^{-10}}{6}$ \\
    Privacy amplification failure probability & $\epsilon_{PA}$ & $\nicefrac{10^{-10}}{6}$ \\
    Error verification failure probability & $\epsilon_{cor}$ & ${10^{-15}}$\\
    Error correction leakage & $f_{EC}$ & 1.16 \\
    & $\lambda_{EC}$ & SI \tnote{a}
    \end{tabular}
        \begin{tablenotes}
        \footnotesize
        \item[a] {see Supplementary Information Sec. \Romannum{6}.E}.
    \end{tablenotes}
    \end{ruledtabular}
    \end{threeparttable}
\end{table}

\subsection{Source performance for in-lab QKD}
We now investigate the performance of the semiconductor SPS for the in-lab QKD experiment. {In Fig. {\ref{fig:lab_qkd}}(a), the SKBs per pulse is shown in dependence of excitation power when the QD is excited under CR of {\SI{76}{\MHz}}.} The average photon number per pulse $\left \langle n \right \rangle$ and blinking-corrected $g^{(2)}(0)$ are measured and fed into Eq. \ref{eq:asymp} and \ref{eq:finite} (see more details in Supplementary Information Sec. \Romannum{6}), assuming a received block size of $n_R^Z=$\SI{e8}{\bits} for the Z-basis in the finite-key regime. Both the asymptotic and finite SKBs per pulse start to drop above the excitation power of \SI{2.98(0.15)}{\micro\watt} because of a decreasing source brightness due to the damping of Rabi-oscillation under p-shell excitation \cite{Ester2008}.

To assess the performance of semiconductor SPS based QKD for long-distance transmission, we study the SKBs per pulse over varying transmission loss in the laboratory, as shown in Fig. \ref{fig:lab_qkd}(b). {The QD is pumped under a CR of {\SI{228}{\MHz}} with the excitation power $\sim$3 times in case of {\SI{76}{\MHz}}} and the emitted single photon signal is coupled into the fibre spools mentioned above. To emulate distances of \SI{20}{\kilo\meter} and \SI{60}{\km}, we added a variable fibre optical attenuator with a fixed loss of \SI{4.0(0.4)}{\dB}. To obtain the data points, the truth table and second-order auto-correlation measurements are recorded based on the statically encoded polarisation qubits at each transmission loss, in order to extract the average photon number per pulse $\left ( \left \langle n \right \rangle \right )$, quantum bit error ratio (QBER) $\left ( e_{tot} \right )$, and single photon purity (see Supplementary Information Sec. \Romannum{6}.B). The solid lines in Fig. \ref{fig:lab_qkd}(b) illustrate the simulation of QBER and SKBs per pulse by employing the values of parameters measured from the QD (Tab. \ref{tab:parameters}). With an increased block size $n_{R}^{Z}$ in the finite-key regime, the simulated maximum tolerable loss (MTL) approaches the one for the asymptotic regime at \SI{28.11}{\dB}, corresponding to a transmission distance of \SI{141.05}{\kilo\meter} in a standard telecommunication fibre. In this experiment, the MTL for both the asymptotic and finite-key block cases are limited by the blinking-corrected $g^{2}(0)$, since the multi-photon emission probability is detrimental for generating high SKRs with single-photon states in the high-loss regime. The SKR and MTL can be improved by employing adequate pre-attenuation \cite{Morrison2023} and employing time gating on the histograms of second-order correlation and truth table during post-processing \cite{Gao2021,Vyvlecka2023}. 

One outstanding feature of SPS is the on-demand photon emission, allowing ultra-high photon count rates with GHz CRs \cite{Schlehahn2016,Shooter2020}. To explore the CR-dependent SKR capabilities of our system, we perform the truth table measurements using \SI{80}{\km} of fibre spool and different excitation laser CRs. {In our experiment, the pump power approximately linearly increases with the CRs based on the saturation power at {\SI{76}{\MHz}} [Fig. {\ref{fig:lab_qkd}}(a)]. However, this is not the case for the {\SI{1063}{\MHz}} because the excitation CR approaches the radiative decay rate of the emitter, as can be seen in the lifetime histograms discussed in the following section.} The QBER, Asymptotic-SKR (A-SKR), and Finite-SKR (F-SKR) are extracted from the truth tables as shown in Tab. \ref{tab:clock_rate}. The QBER decreases with increasing CR due to the lower dark count contributions resulting from smaller integration windows. Although the SKR is limited by the QD lifetime and detector dead times, the achievable high SKRs of $>\SI{100}{\kilo\bits\per\second}$ would enable QKD secured live video conferences encrypted with an one-time-pad (OTP) encryption \cite{Sasaki2011,Liao2018,Zhu2022}. 
\begin{table}[htbp]
    \caption{\label{tab:clock_rate} \textbf{Averaged QBER, A-SKR and F-SKR $\left ( n_R^Z=10^8\right )$  measured under different CRs for the fibre spool distance of \textbf{\SI{80}{\kilo\meter}}}}
    \begin{threeparttable}
    \begin{ruledtabular}
    \begin{tabular}{cccc}
    \textrm{\textbf{CR ({MHz})}}&\textrm{\textbf{QBER (\%)}}&\textrm{\makecell[c]{\textbf{A-SKR (\SI{}{\kilo\bits\per\second})}}}&\textrm{\textbf{F-SKR (\SI{}{\kilo\bits\per\second})}}\\
    \colrule
    76 & 0.344  & 28.12  & 14.19 \\
    228 & 0.099  & 67.97 & 33.88\\
    608 & 0.089  & 151.57 & 75.93\\
    1063 & 0.064  & 216.74 & 108.36\\
    \end{tabular}
    \end{ruledtabular}
    \end{threeparttable}
\end{table}\\

\subsection{Intercity QKD over the deployed fibre}
\begin{figure*}[ht]
    \centering
    \includegraphics[width=\textwidth]{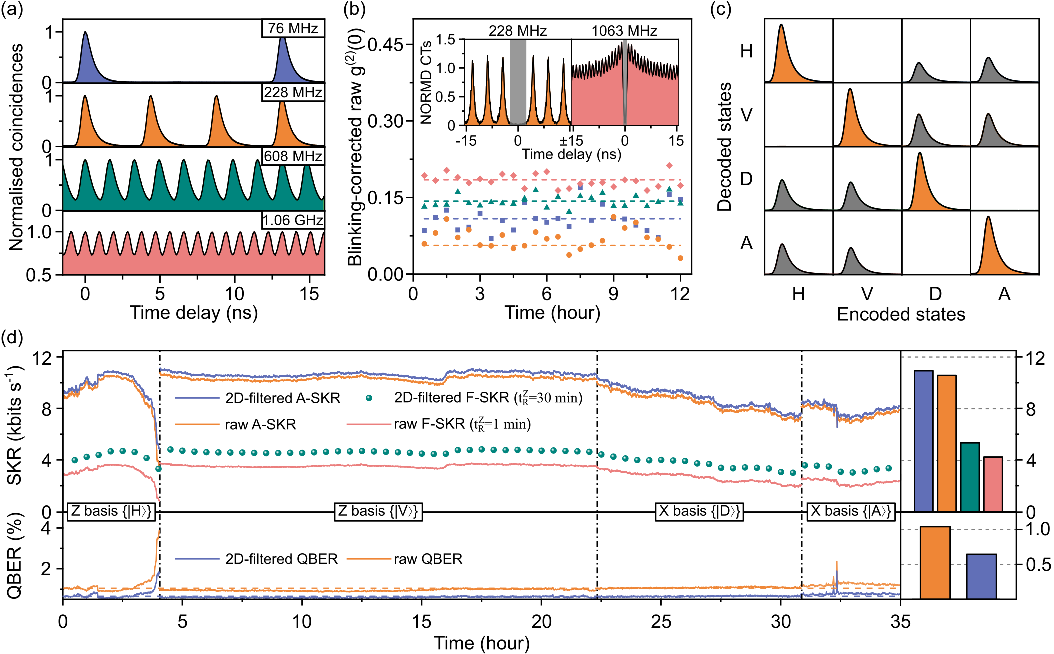}
    \caption{\textbf{Intercity quantum key distribution using telecom C-band single photons from a semiconductor source, transmitted over the deployed fibre {with a loss of {\SI{25.49}{\dB}}.}} (a) Time-resolved QD emission for different CRs. (b) Blinking-corrected, second-order auto-correlation at zero time delay $g^{(2)}(0)$ as a function of measurement time recorded in Braunschweig for the CRs $\textcolor{blue}{\blacksquare}\, \textcolor{blue}{\SI{76}{\MHz}}, \textcolor{orange}{\CIRCLE}\, \textcolor{orange}{\SI{228}{\MHz}},\, \textcolor{green}{\blacktriangle}\, \textcolor{green}{\SI{608}{\MHz}},\, \textrm{and}\, \textcolor{red}{\blacklozenge}\, \textcolor{red}{\SI{1.06}{\GHz}}$. Each data point represents \SI{0.5}{\hour} of measurement time, and no time gating was applied. The dashed lines show the average $g^{(2)}(0)$ over the whole measurement with values of \textcolor{blue}{$0.108 \pm 0.005$}, \textcolor{orange}{$0.056 \pm 0.002$}, \textcolor{green}{$0.143 \pm 0.002$}, and \textcolor{red}{$0.185 \pm 0.002$}. Insets: Normalised (NORMD) second-order correlation histograms for the CRs of \SI{228}{\MHz} and \SI{1064}{\MHz}. (c) Truth table of the polarisation encoded BB84 protocol. The correlation histograms in each column are normalised by the decoded coincidences resulting from the encoded states (diagonal histograms in the table). (d) A-SKR, F-SKR, and QBER over time for different encoding bases $\left \{ X,Z \right \}$ at a fixed CR of \SI{228}{\MHz}. Each data point in the solid lines represents the average of \SI{1}{\minute}. {The colour-coded inset graphs on the right-hand side shows the average asymptotic, finite SKR, as well as QBER over the {\SI{35}{\hour}}- key transmission time with and without the 2-D temporal filter.} The SKBs per pulse are evaluated to be \textcolor{blue}{\SI{4.797(0.011)e-5}{}}, \textcolor{orange}{\SI{4.464(0.011)e-5}{}}, \textcolor{green}{\SI{2.349(0.031)e-5}{}}, \textcolor{red}{\SI{1.867(0.005)e-5}{}}. The average QBER (dashed line) decreases from \textcolor{orange}{\SI{1.041(0.004)}{\percent}} to \textcolor{blue}{\SI{0.646(0.002)}{\percent}}.
    }
    \label{fig:field_qkd}
\end{figure*}

Now, the intercity QKD experiments are performed by sending telecom C-band single photons emitted by the semiconductor QD SPS from Hannover to Braunschweig via the `Niedersachsen Quantum Link'. In the optics laboratory in Braunschweig (Bob), we employ a second SNSPD system (Single Quantum company) to detect the single-photon signals. The detection system's performance in terms of average efficiency, dark count rate has been given in the Section \Romannum{3} of the Supplementary Information. Time traces of the single photon emission under different CRs are obtained by correlating the reference laser and single photon signals [Fig. \ref{fig:field_qkd}(a)]. For CRs ranging from \SI{76}{\MHz} up to \SI{1.06}{\GHz} the single photon pulse trains are clearly identified. Still, at \SI{608}{\MHz} and above the peaks start to overlap, implying a saturation in the achievable photon counts for high CRs. {The Purcell factor of the device can be improved through optimisation of the structure, such as more accurate positioning of QDs {\cite{Rickert2019}} in the cavity centre or modifications in the photonic structure {\cite{Ma2024}}. Alternatively, it can be improved by placing it in an open and tunable micro-cavity within the cryostat {\cite{Tomm2021}}. These methods reduce the radiative lifetime of the QDs towards a higher excitation CR.} Fig. \ref{fig:field_qkd}(b) represents a continuous second-order auto-correlation measurement of up to \SI{12}{\hour} at the deployed fibre end in Braunschweig. The blinking-corrected $g^{(2)}(0)$ plotted over the measurement time reveals high and stable single-photon purity which is important for long-term communication applications. The single-photon purity is preserved at $>\SI{85}{\percent}$ for the first three CRs and reduces to $\sim \SI{81.5}{\percent}$ for \SI{1063}{\MHz} due to coincidence events involving photons from adjacent pulse trains, as visible from the right inset graph. It has to be noted that no temporal filtering has been applied here, which could be used to increase the single-photon purity at higher CRs at the expense of the number of coincidences. The higher blinking-corrected $g^{(2)}(0)$ under \SI{76}{\MHz} in comparison to \SI{228}{\MHz} is due to the contribution of dark counts due to a wider temporal coincidence window (the inverse CR).

To evaluate the performance of real-world QKD over the fibre link, the truth table is measured by accumulating the coincidence histograms between the reference laser and QD signals from the receiver ports, while four polarisation states are statically encoded by the transmitter. An automatic polarisation compensation algorithm at the receiver is developed by minimising the locally measured QBER, in order to counteract polarisation fluctuations of the fibre link (see Supplementary Information Sec. \Romannum{2}.B). Fig. \ref{fig:field_qkd}(c) illustrates the normalised truth table, obtained with a CR of \SI{228}{\MHz} and a measurement time of more than \SI{6}{\hour} for each encoded state. The diagonal histograms in the table correspond to the sifted keys that are usable for error correction and privacy amplification, and the grey histograms illustrate the discarded keys by basis sifting. A fidelity of \SI{99.6}{\percent} is extracted from the projection on the ideal truth table with flawless key decryption. By now measuring the time-dependent truth table, eventually the A-SKR, F-SKR, and QBER time traces are obtained and displayed in Fig. \ref{fig:field_qkd}(d). In addition to extracting the raw time tags, temporal filtering with a 2D-filter is applied by monitoring the $g^{(2)}$ and truth table histograms in order to maximise the SKRs and minimise the QBERs \cite{Gao2021} (see detail in Supplementary Information Sec. \Romannum{7}). The fluctuation of the SKR and QBER while the $\ket{H}$ state is projected onto Z basis, results from the sensitive fibre coupling of $\ket{H}$ signals on the receiver module. {Nevertheless, the dynamic temporal filter reduces the blinking-corrected $g^{(2)}(0)$ from {$\sim$\SI{6.24}{\percent}} to {$\sim$\SI{4.75}{\percent}} with the averaged window size $\sim${\SI{3.56}{\nano\second}} in the full duration of Fig. {\ref{fig:field_qkd}}(d). The averaged QBER down to {\SI{0.646(0.002)}{\percent}} is therefore the lowest value achieved over such a transmission loss so far in QKDs with SPSs, by excluding the {$\sim$\SI{1.17}{\kHz}} noise rate from the raw key rate of {$\sim$\SI{103.16}{\kHz}}. This leads to efficient extraction of secret keys from the X basis, according to the keys sifted by the Z basis for phase error rate estimation.} The average A-SKR and F-SKR are then obtained to be $\SI{10.93(1.19)}{\kilo\bits\per\second}$ and $\SI{5.35(0.58)}{\kilo\bits\per\second}$, respectively. Such SKRs allow for, e.g., live encryption of speech between the two cities via the shared keys \cite{McCree1995}. {Slow polarisation fluctuations are observed on the fibre link (see Supplementary Information Sec. \Romannum{2}.A.3), allowing for effective polarisation drift compensation. For networks in harsh environments where fibres are, e.g., aerially deployed, the time-phase coding protocol {\cite{Takemoto2015}} or time-bin qubits {\cite{Jayakumar2014,Huber2016,Anderson2020,Gines2021}} could be employed with our source, offering less sensitivity to polarisation fluctuations.}

\subsection{Comparison with state-of-the-art}
\begin{figure}[ht]
    \centering
    \includegraphics[width=0.48\textwidth]{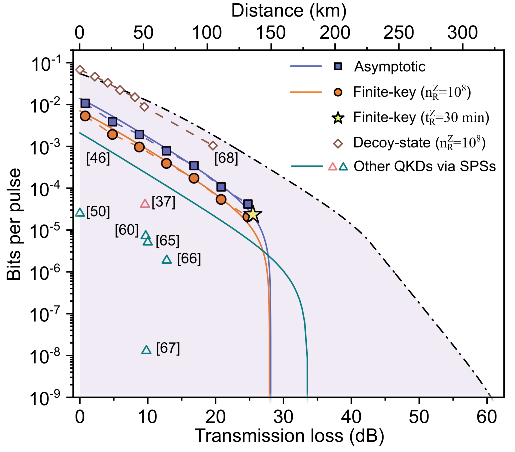}
    \caption{\textbf{Comparison of the SKBs per pulse with other QKD experiments.} The in-lab SKBs per pulse ($\textcolor{blue}{\blacksquare}\, , \, \textcolor{orange}{\CIRCLE}$) and the field-based SKBs per pulse in the finite-key regime ($\textcolor{yellow}{\bigstar}$) surpass previous realisations based on single-photon sources with photons in the telecom C-band and approach state-of-the-art decoy state QKD. The triangles $\left (\textcolor{red}{\triangle}, \textcolor{green}{\triangle} \right )$ show the QKDs with SPSs based via deployed fibre and fibre spools, respectively. The solid lines represent simulated SKBs per pulse using the experimental parameters. The dashed, black line represents the SKBs per pulse that is achievable with experimentally feasible source optimisations. Corresponding distance is calculated for \SI{0.1956}{\dB\per\kilo\meter} standard telecommunication fibre.}
    \label{fig:qkd_comparison}
\end{figure}


Here we present a comparative analysis involving other SPS-based QKD experiments and the high-rate decoy-state BB84 protocol in the telecom C-band as depicted in Fig. \ref{fig:qkd_comparison}. Recent QKD implementations have reported two noteworthy approaches, both utilising SPSs. The first involves QDs in photonic crystal waveguides \cite{Zahidy2023}, while the second employs a QD contained in an oxide-apertured micropillar \cite{Morrison2023}. Both SPS were not directly emitting into the telecom C-band, and therefore frequency conversion techniques were employed, introducing additional losses. In contrast, our implementation features a highly efficient source emitting at the telecom C-band, and both simulation and experimental data reveal a substantial increase in the SKBs per pulse for data transmission. 

Furthermore, we illustrate the asymptotic SKBs per pulse from other QKDs based on telecom single photon emitters \cite{Gao2021,Takemoto2015,Takemoto2010,Intallura2007,Soujaeff2007}. Notably, the achieved finite SKBs per pulse over the intercity fibre testbed, denoted by a star, also represents the highest SKR achieved to date in SPS-based QKD under \SI{228}{\MHz} CR. In addition, our results approach the current record of finite SKBs per pulse achieved by decoy-state BB84 with weak coherent pulses in laboratory settings \cite{Li2023}. 

The demonstrated QKD performance can be further enhanced through optimisations of source quality, detection systems, and protocols. The shaded purple region, delineated by a black dash line, represents an emulation of the achievable finite SKBs per pulse ($n_R^Z=10^8$) by considering experimentally accessible parameters, such as an improved source efficiency \cite{Tomm2021}, single-photon purity \cite{Schweickert2018}, and system dark counts \cite{Liu2023}. Regarding the protocol, optimal pre-attenuation of single-photon counts \cite{Morrison2023} and an asymmetric projection basis choice \cite{Lo2004} for each individual transmission loss result in higher SKR and MTL without compromising security. {A complete, real-time QKD system based on SPSs is considered, which incorporates a high-speed modulation of polarisation states encoded via quantum random numbers. The secret key is then be extracted after error correction and privacy amplification. The additional losses and errors introduced by the additional electro-optical modulator are quantified by using a pre-defined random sequence to encode the polarisation of laser light (see Supplementary Information Sec. \Romannum{8}).} By incorporating these feasible primary parameters (see Supplementary Information Sec. \Romannum{6}.F), we anticipate achieving a MTL of $\sim$\SI{61}{\dB}, corresponding to a distance of \SI{311.86}{\km} in the finite-key regime. Furthermore, a SKR approaching \SI{1}{\MHz} under a transmission loss of approximately $\SI{25}{\dB}$ is attainable assuming a CR of \SI{1}{\GHz}, a configuration well-suited for distributed secure storage \cite{Sasaki2017}. For a more comprehensive comparison of our findings with other QKD protocols, refer to the Supplementary Information Sec. \Romannum{9}.


\section{Discussion}
In conclusion, the first intercity QKD experiments using a deterministic telecom SPS has been demonstrated. This advancement was made possible by harnessing single-photon emissions from a semiconductor QD embedded in a CBG structure, emitting within the telecom C-band and excitation rates up to the GHz range. SKRs have been investigated for different excitation powers and transmission losses under both asymptotic and finite-key scenarios. The measurements and simulations indicate an asymptotic MTL of \SI{28.11}{\dB}, corresponding to \SI{143.71}{\km} channel length in repeaterless quantum communication with standard fibre-optic networks. The experiment spanned a deployed optical fibre link of approximately \SI{79}{\km} and a total loss of \SI{25.49}{\dB}, over which high-rate secret key transmission over an extended period of \SI{35}{\hour} is obtained. The averaged QBER is impressively low at around \SI{0.65}{\percent}, highlighting the robustness and reliability of the system. Comparative analysis with existing QKD systems involving SPS reveals that the SKR achieved in this work goes beyond all current SPS based implementations. Even without further optimisation of the source and setup performance it approaches the levels attained by established decoy state QKD protocols based on weak coherent pulses. This outcome underscores the viability of seamlessly integrating semiconductor single-photon sources into realistic, large-scale and high-capacity quantum communication networks. Moreover, semiconductor QDs, acting as high-speed and deterministic single-photon emitters, hold promising implications for MDI-QKD and may serve as enablers for quantum repeater based star-like networks.

\section{Materials and methods}
\subsection{Source characteristics}
The average photon number per pulse of the linear component of the circularly polarised light from the device is determined using the SNSPD. The p-shell of the QD's trion state was saturably pumped by a pulsed laser at a clock rate (CR) of \SI{76}{\MHz}. The measurement of single-photon count from the transmitter's fibre is conducted over a duration of \SI{1}{\minute} (see Fig. S8 in Supplementary Information \Romannum{4}.A). The average photon count $\mu$ is calibrated at \SI{3.591}{\MHz} with a mean uncertainty of $\nicefrac{\sigma}{\sqrt{N}}=\SI{0.03}{\MHz}$, where $\sigma$ and $N$ represent the standard deviation and the number of data points, respectively. Accounting for the optical transmitter efficiency $\eta_T=$ \SI{46.4(3.4)}{\percent}, SNSPD detection efficiency $\eta_D=$ \SI{74.0(6.0)}{\percent}, and CR of \SI{76}{\MHz}, the average number of photons per pulse for the linearly polarised single photons is calibrated to be $\left\langle n \right\rangle$ =\SI{0.138(0.015)}{}.

In our experiments, we excited the quantum dot in a quasi-resonant manner. The exciton is initially excited to the p-shell of the QD. Subsequently, it decays to the s-shell and, thereafter to the ground state by emitting a single photon. Hence to fit the lifetime, we assume, that a decay model for a three-level system is sufficiently accurate (see Fig. S9 in Supplementary Information Sec. \Romannum{4}.B). These then lead to the following identity, which is used to describe the population in the s-shell:
\begin{align}
f(t) = a_0  \cdot   \frac{T_{s}}{ T_{p} -  T_{s} } \cdot \left [ \exp \left ( - \frac{t - t_0}{ T_{p}} \right )  - \exp \left ( - \frac{t - t_0}{ T_{s}} \right )  \right ]
\label{eq:lifetime}
\end{align}
\begin{align}
	g(t) = \frac{1}{ \sigma \sqrt{2 \pi} } \cdot \exp \left (\frac{-t^2}{2\sigma^2} \right )
\label{eq:gaussian}
\end{align}
\begin{align}
	N_s(t) =  (f * g)(t)
	\label{equ:lifetime_extraction}
\end{align}

In order to account for the time jitter of the detector and the mode shape of the laser pulse, we employ least-square iterative re-convolution with the instrument response function, which is fitted with a Gaussian function (Eq. \ref{eq:gaussian}). The lifetimes of $T_{p}$ and $T_{s}$ are extracted as \SI{149.3(1.0)}{\pico\second} and \SI{443.20(1.58)}{\pico\second} using Eq. \ref{equ:lifetime_extraction}, yielding the total lifetime of \SI{592.5(1.8)}{\pico\second} with the cascade process.

In SPS-based QKD, the average photon number per pulse and single-photon purity are two essential parameters that must be considered in the QKD algorithm to calculate the secret key rate and the maximum tolerated loss. For the non-blinking SPS, the single-photon purity is typically assessed as $\left [ 1-g^{(2)}(0) \right ]$ \cite{Vyvlecka2023}. However, the regular definition of $g^{(2)}(0)$ with blinking effect will be higher than the non-blinking $g^{(2)}(0)$, and this results in the underestimation of the single photon purity with $\left [ 1-g^{(2)}(0) \right ]$. To analyse the asymptotic and finite secret key rates in our experiment, we employed the blinking-corrected $g^{(2)}(0)$ \cite{Schimpf2021}. We begin by integrating the raw second-order auto-correlation histogram with a temporal bin size of approximately $\SI{4.38}{\nano\second}$ (the inverted value of the \SI{228}{\MHz} pulsed-CR). The raw $g^{(2)}(0)$ value without blinking correction and temporal filtering on the background is \SI{2.95(0.02)}{\percent}. To correct the blinking, we apply the blinking fitting function by calculating the ratio between the measured data and fitted value \cite{Schimpf2021}. The normalised second-order auto-correlation after the blinking correction is then obtained, from which we extract the blinking-corrected $g^{(2)}(0)=$\SI{2.43(0.02)}{\percent} (see more details in Supplementary Information Sec. \Romannum{4}.C).

\subsection{Experimental setups}
The experimental setups includes the transmitter, fibre spools, receiver, and the SNSPD (see Fig. S1 in Supplementary Information Sec. \Romannum{1}). In the transmitter, the clock variable pulsed fibre laser is coupled into free space to excite the QD after passing through a beam splitter (Altechna company) with a splitting ratio of (R:T=98.5:1.5). The Attodry1100 system is equipped with a Thorlabs aspheric lens (C330TMD-C) with a numerical aperture of 0.7 to collected the single photons from the device. Thorlabs polariser (LPNIR050-MP2), zero-order half-waveplate (WPH05M-1550), and quarter-waveplate (WPQ05M-1550) are employed to purify and encode the polarisation states, respectively. The Thorlabs electronic stages (K10CR1/M) control these components. The encoded single photon qubits are coupled into Corning SMF-28® Ultra fibre spools, each spanning \SI{40}{\km}. The receiver utilises a fibre-based Bragg grating (\SI{1560}{\nano\meter} half-wave CWDM) to split the reference laser and single-photon signals. The single photons, which are encoded, are detected by the SNSPDs (Single Quantum company) after passing through a 50:50 beam splitter (10B20NP.31) and two plate polarising beam splitters (PBSW-1550). Finally, the photon arrival times are registered by the timetagger (Time Tagger Ultra from Swabian Instrument company). More details about the efficiencies of the transmitter and receiver are shown in the section \Romannum{1}.A and \Romannum{1}.B of the Supplementary Information.

\section*{Data availability}
The data that support the plots within this paper and other findings of this study are available from the corresponding author upon reasonable request.

\section*{Acknowledgments}
We thank J. Wang, C. Nawrath, M. Auer, T. van Leent and H. Weinfurter for fruitful discussions, R. Guan for helping with the optical setups, and J. Wang for the electronics control of the system. We thank J. Kronj\"{a}ger, A.K. Kniggendorf and A. Kuhl for efficiently organising the deployed fibre infra-structure. We would also like to thank the companies Single Quantum, Quantum Optics Jena, and PriTel Inc. for their continued and timely support.

The authors gratefully acknowledge the funding by the German Federal Ministry of Education and Research (BMBF) within the project QR.X (16KISQ013 and 16KISQ015), SQuaD (16KISQ117) and SemIQON (13N16291), the European Research Council (QD-NOMS GA715770, MiNet GA101043851), European Union’s Horizon 2020 research and innovation program under Grant Agreement No. 899814 (Qurope), EMPIR programme co-financed by the Participating States and from the European Union’s Horizon 2020 research and innovation programme (20FUN05 SEQUME), the Deutsche Forschungsgemeinschaft (DFG, German Research Foundation) within the project InterSync (GZ: INST 187/880-1 AOBJ: 683478), and under Germany’s Excellence Strategy (EXC-2123) Quantum Frontiers (390837967), and Flexible Funds programme by Leibniz University Hannover (51410122).

\section*{Conflict of interest}
The authors declare no competing interests.

\section*{Contributions}
J. Yang and R. Joos implemented the optical characterisation of the sample. J. Yang, Z. Jiang, F. Benthin and J. Hanel, A. Hreibi and M. Zopf carried out the QKD experiment. S. Bauer, S. Kolatschek, and M. Jetter designed and fabricated the sample. T. Fandrich, J. Hanel, J. Yang, F. Benthin and Z. Jiang performed the data analysis and the interpretation of the results. J. Yang wrote the manuscript with the help from M. Zopf, F. Ding, E. P. Rugeramigabo, S.L. Portalupi, P. Michler, and the other co-authors. F. Ding, P. Michler, and S. K\"{u}ck conceived and supervised the project.

\nocite{}
\bibliography{reference}
\end{document}